# Ultrafast switching of sliding ferroelectricity and dynamical magnetic field in van der Waals bilayer induced by light

Jian Wang[1†], Xu Li[1†], Xingyue Ma[1], Lan Chen[1,2], Jun-Ming Liu[1], Chun-Gang Duan[3,4], Jorge Íñiguez-González[5,6*], Di Wu[1*], Yurong Yang[1*]

[1]*Laboratory of Solid State Microstructures, Jiangsu Key Laboratory of Artificial Functional Materials, Nanjing University, Nanjing 210093, China*

[2]*School of Science, Nanjing University of Posts and Telecommunications, Nanjing, 210023, China*

[3]*Key Laboratory of Polar Materials and Devices (MOE) and State Key Laboratory of Precision Spectroscopy, East China Normal University, 500 Dongchuan Rd., Shanghai 200241, China*

[4]*Collaborative Innovation Center of Extreme Optics, Shanxi University, Taiyuan 237016, China*

[5]*Department of Materials Research and Technology, Luxembourg Institute of Science and Technology, 5 Avenue des Hauts-Fourneaux, L-4362 Esch/Alzette, Luxembourg*

[6]*Department of Physics and Materials Science, University of Luxembourg, 41 Rue du Brill, L-4422 Belvaux, Luxembourg*

**Sliding ferroelectricity is a unique type of polarity recently observed in a properly stacked van der Waals bilayer. However, electric-field control of sliding ferroelectricity is hard and could induce large coercive electric fields and serious leakage currents which corrode the ferroelectricity and electronic properties, which are essential for modern two-dimensional electronics and optoelectronics. Here, we proposed laser-pulse deterministic control of sliding ferroelectricity in bilayer *h*-BN by first principles and molecular dynamics simulation with machine-learned force fields. The laser pulses excite shear modes which exhibit certain directional movements of lateral sliding between bilayers. The vibration of excited modes under laser pulses is predicted to overcome the energy barrier and achieve the switching of sliding ferroelectricity. Furthermore, it is found that three possible sliding transitions – between AB (BA) and BA (AB) stacking – can lead to the occurrence of dynamical magnetic fields along three different directions. Remarkably, the magnetic fields are generated by the simple linear motion of**

---

† These authors contributed equally to this work.
*Corresponding author. Email: jorge.iniguez@list.lu (J.I.); diwu@nju.edu.cn (D.W.); yangyr@nju.edu.cn (Y.Y.)

**nonmagnetic species, without any need for more exotic (circular, spiral) pathways. Such predictions of deterministic control of sliding ferroelectricity and multi-states of dynamical magnetic field thus expand the potential applications of sliding ferroelectricity in memory and electronic devices.**

# Introduction

Ferroelectrics are important functional materials with switchable electric polarization that have been widely used as sensors, memories, actuators, and other electronic and electro-optic devices[1-6]. In the quest for greater integration, low dimension, and small size in modern electronics, two-dimensional ferroelectrics with a few atomic layers are proposed and synthesized[7-9], such as $CuInP_2S_6$[10, 11], $SnTe$[12], $In_2Se_3$[13-15], $BA_2PbCl_4$[16], and $NiI_2$[17, 18]. Exotic properties such as negative piezoelectricity[19] have been revealed in these systems[20, 21]. Very recently, a unique type of polarity of sliding ferroelectricity was reported that arises from a properly stacked van der Waals bilayer where the spatial inversion symmetry is broken and the polarization can be switched by sliding one monolayer with respect to another (see Fig. 1). The sliding ferroelectricity was first proposed in bilayers of $h$-BN and $MoS_2$ by Li and Wu[22]. The sliding ferroelectricity in bilayer $h$-BN was then firstly experimentally observed[23, 24]. Other sliding ferroelectric multilayers such as $WTe_2$[25-27], $MoS_2$[28, 29], and other metal dichalcogenides layers[30] were also observed. Systematic group theory was used to analyze sliding ferroelectricity and a general rule to govern the generation of sliding ferroelectricity was proposed[31]. Despite these achievements, the direct observation of pure sliding ferroelectricity by polarization–electric field hysteresis has not been reported until now[32], and most of the ferroelectric properties were characterized by transport behaviors or piezo-response force microscopy (PFM)[23, 33]. The fact that the driving electric field needs to be applied perpendicular to the plane of the bilayer, while the sliding movement is within the plane, further complicates the realization of ferroelectric switching. Furthermore, the electric-field-driven switching of ferroelectricity is slow (at the timescale of nanoseconds or slower)[4, 34]. Therefore, it is highly required to find an alternative driving force to switch the polarization of sliding ferroelectrics.

Light is another approach to switch ferroelectric polarization which has been proposed and achieved in perovskite oxides experimentally recently[35-38]. It was demonstrated that the polarization in ferroelectric perovskites can be reversed by exciting a high-frequency infrared-active phonon mode that couples to the soft mode (which leads to the polarization), with the effect acting through an intermediate anharmonic driving force[36, 37, 39, 40]. It is unclear whether a similar strategy could be applied to van der Waals layers displaying sliding polarization, as the phonon-mediated couplings between layers can be expected to be small. Hence, this remains an open question.

Switching ferroelectric polarization in perovskites could induce the dynamical magnetic moment through $\boldsymbol{M} \propto \boldsymbol{P} \times \partial_t \boldsymbol{P}$, where $\boldsymbol{M}$ is the magnetic moment and $\boldsymbol{P}$ is the polarization. Typically, having a non-zero cross-product between $\boldsymbol{P}$ and $\partial_t \boldsymbol{P}$ requires ions to move in a circular or spiral pathway and carry an angular momentum. Interestingly, in the switching of bilayer stacking ferroelectricity, the ions move within the plane while affecting the out-of-plane polarization (see Fig. 1). It is therefore legitimately to wonder if the dynamical magnetic moment can be generated in the switching of sliding ferroelectricity.

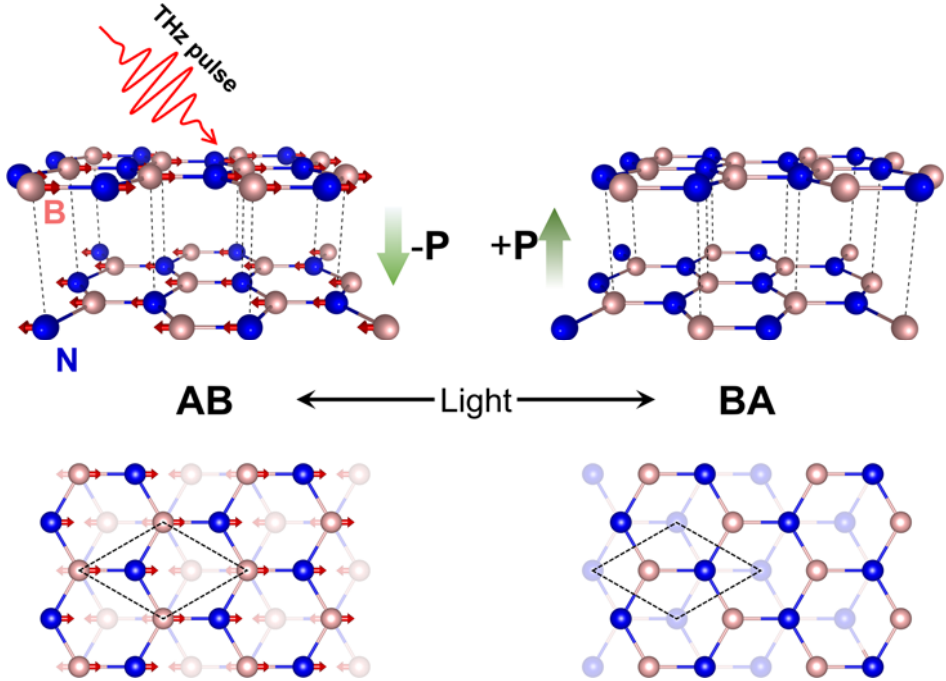

**Fig. 1. Schematic of the transition between the AB and BA stacking of bilayer *h*-BN by the excited shear phonon under laser pulses.** Top view (top panels) and side view (bottom panels) of bilayer *h*-BN. Pink and blue spheres denote the B and N atoms, respectively. The red arrows represent the eigenvector of the shear phonon mode. Dash lines denote the correlation of atoms in the vertical direction. The green arrows represent the direction of polarization.

Here, we studied the switching of sliding ferroelectricity in bilayers $h$-BN and the induced dynamical magnetic moment under terahertz (THz) pulsed laser using first principles and molecular dynamics with a machine learning force field. We find that the laser pulses excite the shear modes and drive the switching of sliding ferroelectricity within a few picoseconds. We also show that this ultrafast switching of sliding ferroelectricity can further induce dynamical magnetic moment, though the atoms of bilayers move linearly (rather than spirally as in perovskites[41-43]). This study thus proposes a novel approach to the control of the sliding ferroelectricity and the induced dynamical magnetic field further expanding the potential application of sliding ferroelectricity of van der Waals materials.

**Results**

**Light-induced switching of sliding ferroelectricity**

Bilayer $h$-BN possesses AB stacking or BA stacking, as shown in Fig. 1. In AB (BA) stacking the B (N) atoms in the top layer sit above N (B) atoms in the bottom layer while N (B) atoms in the top layer are above the empty site of the center of the hexagon in the bottom layer. The asymmetric $p_z$ orbitals distribution of B and N give rise to a net electric dipole and thus a polarization[23, 44]. As a result, AB and BA stacking exhibit out-of-plane polarization in opposite directions (Fig. 1), and sliding bilayer between the two stacking could achieve the switching of out-of-plane polarization.

Figure 2 displays the phonon dispersion of the bilayer $h$-BN. Three acoustic modes linearly increase from zero frequency near Γ point. There are three optical $A_1$ modes and six optical $E$ modes (see Supplementary Table 1). The lowest energy optical modes labeled $E(4)$ and $E(5)$ are degenerate at Γ point with a frequency of 1.085 THz. These two modes are shear modes, where the atoms in one layer all move in one direction while the atoms in the other layer move in the opposite direction. As shown in Fig. 2b, for $E(4)$ mode the top layer moves in -$y$ direction (parallel to the armchair direction of $h$-BN) while the bottom layer moves in $y$ direction. The displacements of $E(5)$ mode are along the $x$/-$x$ direction (parallel to the zigzag direction of $h$-BN), perpendicular to that of $E(4)$. The vibration of the shear modes of $E(4)$ and $E(5)$ exhibits the behavior of sliding the top layer relative to the bottom layer, which may induce the switching of out-of-plane polarization between stacking patterns AB and BA.

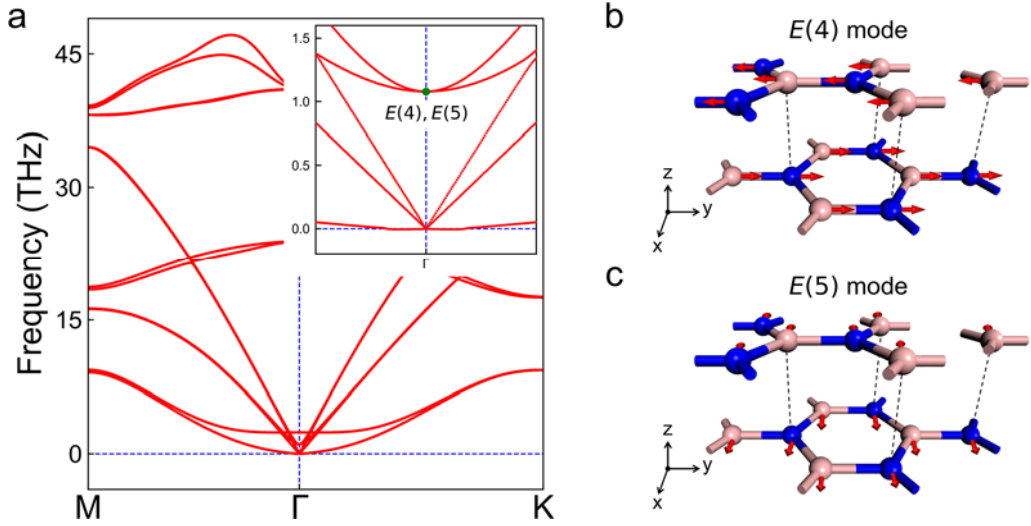

**Fig. 2. The phonon dispersion and shear normal modes of bilayer *h*-BN. a** The phonon spectrum of AB-stacked *h*-BN. The inset in panel (a) shows the zoomed low-frequency dispersion near Γ point. The green dot in the inset denotes the energy of degenerate and orthogonal phonon modes of $E(4)$ and $E(5)$. The eigenvectors of the (**b**) $E(4)$ mode and (**c**) $E(5)$ mode.

Figure 3a displays the total energy change when the top layer slides in the two-dimensional plane. One can clearly see that the AA stacking has the maximum energy and that the ferroelectric switching pathway corresponds to a straightforward sliding from the AB to the BA stacking, or vice versa. Figure 3b displays the energy change when sliding the bilayer from AB to BA stacking. We use the amplitude of the $E(4)$ mode ($Q[E(4)]$) to describe the sliding movement. $Q[E(4)]$ is zero for the AB stacking and 5.1 Å√amu at BA stacking. One can confirm that we have an energy barrier of 2.2 meV/f.u. for switching this sliding ferroelectric polarization $P_z$ (see Fig. 3c, the maximum $P_z$ is 1.43 pC/m). The structure of the lowest energy barrier for the switching lies at the saddle-point (SP) where $Q[E(4)]$=2.55 Å√amu. Note that there are three energy-degenerate BA structures to which one AB structure can be translated, due to the $C_3$ symmetry of the system, as shown in Fig. 3a, indicating three possible low-energy ferroelectric switching pathways.

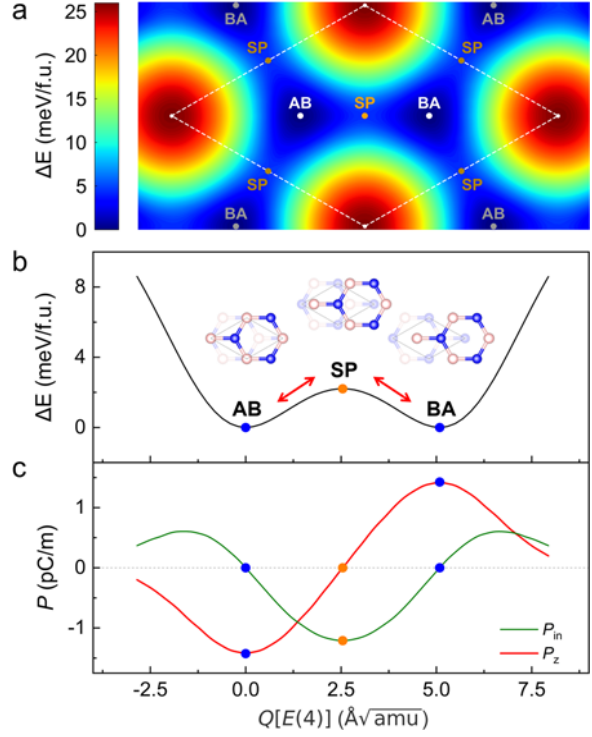

**Fig. 3. Physical quantities with respect to the amplitude of *E*(4) mode in bilayer *h*-BN. a** The energy surface of sliding one layer respective to the other layer. The energy (**b**) and polarizations (**c**) as functions of the amplitude of *E*(4) mode are displayed. The blue spots marked with AB (BA) represent the corresponding stacking configurations, and the orange spot denotes the saddle-point (SP) barrier during the sliding transition. The insets in panel **b** show the top view structures at AB, SP, and BA points in the pathway. In-plane polarization is along the *y* direction, parallel to the sliding direction from BA to AB point.

Light with terahertz frequency can be used to excite phonons. We mimic the pulsed lasers by applying the Gaussian-enveloped electric field in the form of $\mathbf{E} = \mathbf{E}_0 \exp(-4\ln 2\, t^2/\sigma^2) \cdot \cos(2\pi\omega t)$, where $\mathbf{E}_0$ represents the peak of the electric field while $\omega$ and $\sigma$ refer to the frequency and full width at half maximum (FWHM) of the pulse, respectively. The application of pulsed lasers is simulated using molecular dynamics with the aid of a machine-learned force field in an 8×8 supercell (see Methods). Here, we do not consider the photoelectron excitation in the simulations because of the significantly large band gap of about 6 eV for *h*-BN[45], which is much larger than phonon energies and the frequency $\omega$ of the considered pulsed laser. First, a single pulse with 1 THz frequency (thus, very close to the frequency of the *E*(4) and *E*(5) modes displaying the sliding behavior) and polarized direction along *y* is used at a temperature of 50 K. The vibration of bilayer *h*-BN under electric pulses could be projected into the amplitudes of 9 optical modes at the Brillouin zone center of the unit cell. Figure 4a shows the evolution of the amplitude of 9 optical modes when applying a single pulse. The magnitude Q for the *E*(4) and *E*(5) modes are much larger than the other optical modes, indicating these two modes are resonant and activated by pulses. The *E*(4) mode with a vibration direction parallel/antiparallel to the electric pulse is much more strongly activated than the *E*(5) whose vibration direction is perpendicular to the pulse. One can thus determine the sliding direction of bilayer *h*-BN by selecting the polarized direction of the pulses. Besides, the amplitude of *E*(4) can be further enhanced by increasing the peak electric field (Supplementary Fig. 2).

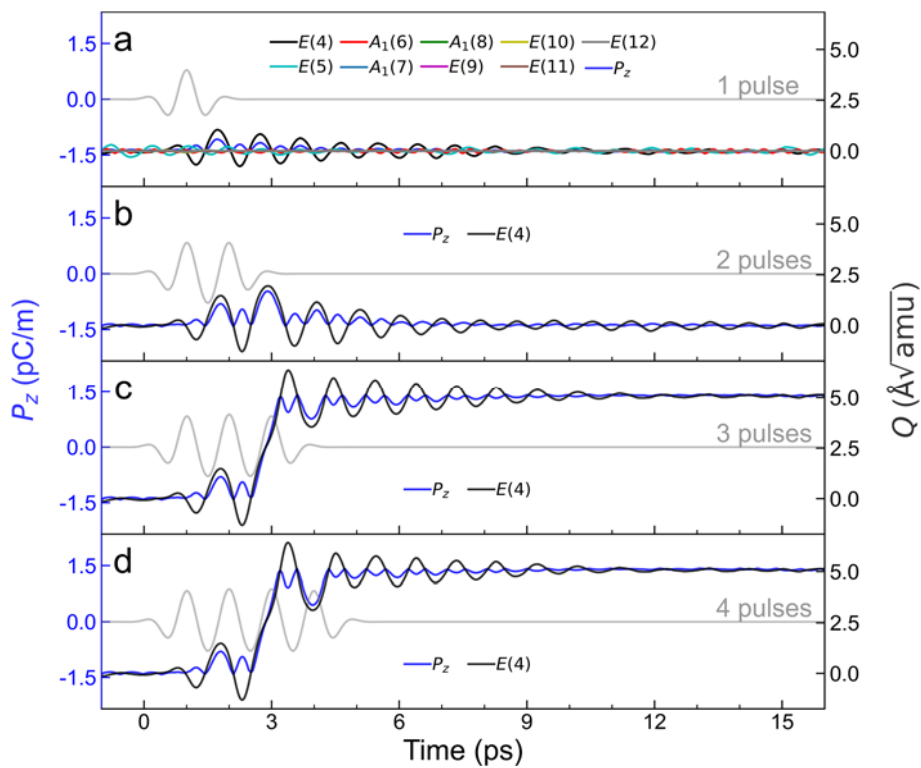

**Fig. 4. Time evolution of out-of-plane polarization and amplitude of phonon modes. a** Out-of-plane polarization and the optical mode decomposition of the vibration as functions of time under a single pulse. The polarization and amplitude of mode *E*(4) as functions of time under (**b**) two pulses, (**c**) three pulses, and (**d**) four pulses. The grey lines represent the electrical-field form of pulses, with a frequency of 1 THz. The train of pulses is separated by 1 ps. The maximum amplitude of the electric field used here is 0.4 V/Å. The simulations are performed at 50 K.

We then consider situations where the bilayer $h$-BN is under multi-pulsed lasers with a specified interval between each pulse (see Methods). Figure 4b-d shows the evolution of quantities under two pulses, three pulses, and four pulses, respectively, at 50 K. Note that the train of pulses is separated by 1 ps. The activated maximum Q of $E(4)$ under two pulses is 1.95 Å$\sqrt{\text{amu}}$, about twice larger than that of 1.04 Å$\sqrt{\text{amu}}$ under a single pulse. The maximum of Q is delayed by about 1 ps compared to the second maximum peak of the pulse. The activated Q slightly decays when laser pulses are stopped and fully annihilates in the thermal vibration after 11ps. When further increasing trains of pulses to three and four pulses, Q of $E(4)$ further increases and overcomes the energy barrier at the SP (see Fig. 3) and finally increases to about 5.1 Å$\sqrt{\text{amu}}$, resulting in the transition from AB stacking to BA stacking and the switching of out-of-plane polarization $P_z$ (see the movements in SupplementaryMovie1.gif). This switching of sliding ferroelectricity by THz pulses is achieved within 3-4 ps, much faster than the switching by an electric field which is achieved within the timescale of nanosecond[4, 34].

**Dynamical magnetic field**

AB or BA stacked structure with a layer group of $p3m1$ has an out-of-plane polarization of 1.43 pC/m and zero in-plane polarization. By contrast, the structures in the switching pathway from AB to BA (and vice versa) present the layer group $cm11$, exhibiting both out-of-plane and in-plane polarization components. One exception is the structure at the SP point where the out-of-plane polarization vanishes and the in-plane polarization reaches the maximum, as shown in Fig. 3b (also Fig. 5). Hence, through the switching path, the polarization evolves in a way that resembles the cycloid-type modulation, typical of a 180º Néel-type domain wall[46] (see Fig. 5a). This cycloid-like dynamical rotation of polarization could induce dynamical magnetic field[43]. This magnetic moment **M** of a unit cell is given by (more details in Supplementary Information)

$$\mathbf{M} = \frac{S^2 m_e}{e^2 \hbar} \frac{\bar{Z}^*}{\bar{Z}^*_{yy} \bar{Z}^*_{zz}} \mathbf{P} \times \partial_t \mathbf{P}, \tag{1}$$

where $S$ is the area of 2D $h$-BN per unit cell, $m_e$ is the electron mass, $e$ is the elementary charge, and **P** is the polarization vector. $\bar{Z}^*_{xx}$, $\bar{Z}^*_{yy}$ and $\bar{Z}^*_{zz}$ refer to the *xx*, *yy*, and *zz* components of the Born effective charge tensor, respectively. $\bar{Z}^*$ is the average of $\bar{Z}^*_{yy}$, $\bar{Z}^*_{zz}$ and $\bar{Z}^*_{xx}$. The corresponding dynamical internal magnetic field **B** is then given by $\mathbf{B} = \mu_0 \mathbf{M}/V$, where $\mu_0$ is the vacuum permeability and $V$ is the volume of the unit cell.

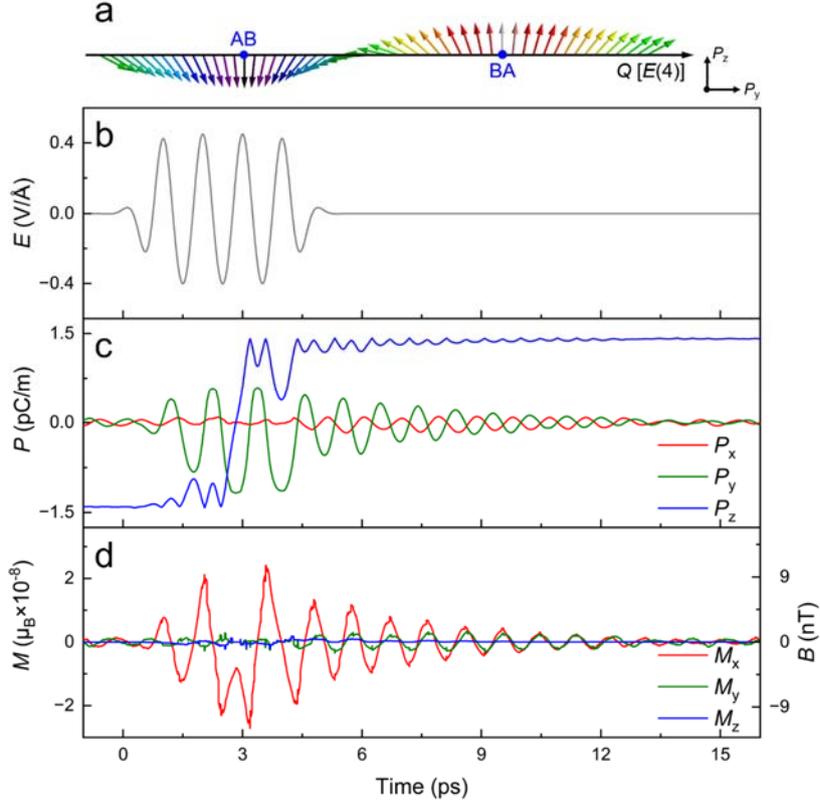

**Fig. 5. Time evolution of polarization and dynamic magnetic moment under laser pulses. a** The sketch of the polarization vector changes when sliding the bilayer between AB and BA stackings. **b** Electric field pulses used in simulations leading to the time-dependent polarization components ($P_x$, $P_y$, and $P_z$) shown in panel (**c**) and the time-dependent magnetic moment components ($M_x$, $M_y$, and $M_z$) in panel (**d**). The *x*, *y*, and *z* directions correspond to those defined in Fig. 2.

Figure 5c shows the time evolution of polarization under the train of pulses shown in Fig. 5b at a temperature of 5 K. In the dynamical switching process, out-of-plane polarization changes from -$P_z$ to +$P_z$ at about 3 ps, and a dynamical in-plane polarization along the *y* direction is induced during the time of the pulsed laser. For the in-plane polarization, the negative magnitude is larger than the positive magnitude, indicating a net polarization along the -*y* direction, consistent with the static illustration shown in Fig. 3b. By Eq. (1), we calculated the dynamical magnetic moment in the switching process. Figure 5d shows the dynamical magnetic moment components of $M_x$, $M_y$, and $M_z$ along the $x$, $y$, and $z$ directions, respectively. $M_y$ and $M_z$ are zero, while $M_x$ reaches a maximum magnitude of $2.7\times10^{-8}$ $\mu_B$ and decays along with the decaying of $P_y$ when laser pulses stop. The corresponding maximal internal dynamical magnetic field is about 12 nT, which lies in the range of experimentally achievable sensitivities using NV center magnetometry[47, 48]. Note that the magnitude of dynamical magnetic moment (magnetic field) could be increased by more than one order of magnitude in two-dimensional materials with a larger sliding ferroelectric polarization such as ZnO, GaN[22, 33]. It is very interesting to note that this dynamical magnetic field is induced by simple linear ionic movements, at variance from the orbital, spiral, and other complex pathways proposed before[42, 43]. Also note that the polarization rotates in cycloid-type in bilayer *h*-BN and only an in-plane magnetic field is induced. If the polarization rotates helically such as that in CrI$_3$[31], the out-of-plane magnetic field could also be generated. This dynamical magnetic field may interact with the intrinsic magnetism of magnetic bilayer (such as the antiferromagnetism of CrI$_3$ bilayers[49] or ferromagnetism in Cr$_2$Ge$_2$Te$_6$[50]) and thus may yield novel phenomena such as Zeeman splitting[42], electromagnon[51], and topological magnetic configurations[52].

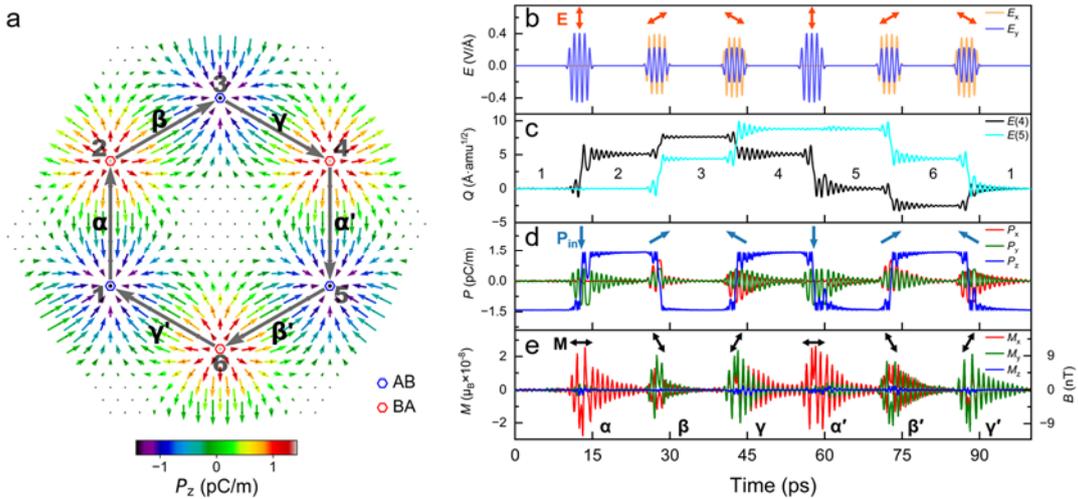

**Fig. 6. Dynamical magneto-electric effect for sliding the bilayer in a hexagonal pathway. a** The polarization vector for sliding bilayer in two-dimensional plane. The states of 1, 3, and 5 are AB stacking, and the states of 2, 4, and 6 are BA stacking. The in-plane and out-of-plane polarization components are represented by the short arrows and their colors, respectively. The long grey arrows represent the sliding direction. The time evolution of (**b**) electric field, (**c**) amplitude of *E*(4) and *E*(5) modes, (**d**) electrical polarization, and (**e**) magnetic moment. The numbers in panel **c** denote the states corresponding to the states shown in panel **a**. The Greek letters α, β, γ, α′, β′, and γ′ in panel **e** denote the dynamic states corresponding to those in panel **a**. The orange arrows in panel **b**, blue arrows in panel **d**, and black arrows in panel **e** represent the directions of the electric field, in-plane polarization, and magnetic moment, respectively.

As mentioned above, the bilayer *h*-BN has a three-fold symmetry and, thus, a given AB stacking structure can switch into any of three equivalent BA structures. The three possible sliding directions make angles of 120º with each other (see inset of Fig. 3a). The bilayer sliding direction can be determined by the polarized direction of the laser pulses, which resonate with a particular combination of the degenerate *E*(4) and *E*(5) sliding modes. Therefore, by modulating the polarized direction of the pulses, it is possible to create a six-fold switching pathway between the six stacking structures 1→2→3→4→5→6→1, as shown in Fig. 6a (see the movements in SupplementaryMovie2.gif). States 1, 3, and 5 are AB stacking and states 2, 4, and 6 are BA stacking. Figure 6b shows the six trains of pulses, with a 10 ps interval of zero field between the trains of pulses. The polarized directions of pulses are shown in the inset of Fig. 6b, which are along six sides of a hexagon (see Fig. 6a). The displacements of each sliding can be represented by amplitudes of phonon modes *E*(4) along the *y* direction and *E*(5) along the *x* direction, respectively.

As shown in Figs. 6c and 6d, the amplitude of modes and out-of-plane polarization ($P_z$) for the six states are stable in the separation between the train of pulses. Figures 6d and 6e show the dynamical in-plane polarization and dynamical magnetic moment. The in-plane polarization ($\mathbf{P}_{in}$) is induced in the switching process and decays at the periods between pulse trains. The dynamical $\mathbf{P}_{in}$ rotates its direction anticlockwise by 120º in each switching from state *n* to state *n*+1 (*n*=1~6, 7 represents to 1) (see inset arrows in Fig. 6d). The out-of-plane magnetic moment ($M_z$) is zero, implying that the magnetic vector **M** is in the plane (see Fig. 6e). This dynamical **M** reaches a maximum magnitude of about $2\times10^{-8}$ $\mu_B$, corresponding to the magnetic field of a few nT which can be detected by NV center magnetometry[47, 48]. The magnetic moment decays towards zero after each pulse train. The dynamical **M** rotates its direction by 60º clockwise in each switching from state *n* to state *n*+1 (*n*=1~6, 7 represents to 1) (see inset arrows in Fig. 6e) in the six-fold sliding route shown in Fig. 6a. Therefore, thanks to the sliding ferroelectric switching, one can induce 3 types of dynamical magnetic fields with similar amplitudes and along three different in-plane directions.

**Perspective**

In the switching of sliding ferroelectricity in the bilayer *h*-BN, the train with 3 or 4 pulses is enough to achieve the switching. Note that, for the switching to be deterministic, it is critical to control the number of pulses. Indeed, if the pulses are continuously used without separation, the bilayer will switch back to the original state at the time of the 10th pulse, approximately, and the bilayer would jump between the original state and the switched one. Let us also stress that, compared to the switching of ferroelectricity in perovskites, there are no domain nucleation-and-growth kinetics or related sources of disorder in the switching of sliding ferroelectricity, due to the sliding movement in the whole sample. Therefore, this constitutes a unique example of monodomain ferroelectric switching, potentially resilient to fatigue problems, which makes the sliding ferroelectric bilayer very promising for nanodevices.

Every AB (BA) stacking structure can switch to any one of three symmetry-equivalent BA (AB) states, depending on the polarized direction of the applied pulses. Accordingly, three differently oriented dynamical magnetic fields can be obtained, each of them with a different phase. As shown in Supplementary Fig. 5, we symbolize these dynamical magnetic fields by α (α′), β (β′), and γ (γ′), as those in Figs. 6a and 6e. The switching from an AB stacking (the center of the hexagon of Supplementary Fig. 5) with -$P_z$ to a BA stacking with +$P_z$ could lead to one of α, β′, and γ dynamical magnetic fields. Conversely, the switching from a BA stacking (e.g., the top of the hexagon of Supplementary Fig. 5) with +$P_z$ into a stacking with -$P_z$ could lead to a dynamical magnetic field of α′, β, and γ′. These possible dynamical signals induced in each light-driven switching of sliding ferroelectricity suggest a novel dynamic magnetoelectricity at an ultrafast speed. Additionally, this novel dynamic magnetoelectricity allows conversion from charge (polarization) to spin with controlled magnetization direction, suggesting a magneto-optical memory that can be written by laser pulses and read magnetically at an ultrafast speed. These effects of light-polarization-magnetization in van der Waals magnetic materials may be further used to detect Zeeman splitting[53], electromagnon[54], and spin-density waves[55, 56].

In summary, in this work, we propose a laser-pulse-driven ultrafast switching of sliding ferroelectricity, as predicted by first principles and molecular dynamics simulation using machine-learned force fields. The ultrafast switching further induces dynamical magnetic fields with three possible directions corresponding to the possible interlayer sliding directions. The light control of ultrafast switching and the induced multi-states of dynamical magnetic field implies the potential application for ultrafast memory and spintronic-related devices, e.g., to achieve charge-to-spin conversion.

# Methods

## Computational details.

The first-principles calculations are performed based on Vienna ab initio simulation package (VASP) code[57, 58]. PBEsol function[59] based on projector-augmented wave method[60] is used. We use the kinetic energy cutoff of 550 eV and the total energy convergence criterion of $10^{-8}$ eV. The Brillouin zone is sampled with at least 21×21×1 Monkhorst-Pack k-mesh for primitive cells. For geometric optimization, the forces on all atoms are optimized to less than $5\times10^{-4}$ eV/Å. A vacuum space with a vertical distance greater than 20 Å is constructed and the PBEsol-D3 method of Grimme with Becke-Jonson damping is used to take into account the van der Waals interaction[61]. The polarization is calculated by the Berry phase method[62] and the product of the atomic displacements with the Born effective charges[63].

## Simulations of applying pulsed lasers.

The molecular dynamics (MD) simulations are performed with the machine learning force field (MLFF)[64-66]. The root mean square errors of the MLFFs from the training data sets are less than 0.25 meV/atom in energy, 0.03 eV/Å in force, and 0.23 kbar in stress tensors. These roots mean square errors are similar to those in previous MLFF simulations[64, 67, 68]. A supercell with 256 atoms is used within the NpT ensemble using a Langevin thermostat[69] and the time step of 1 fs. The laser pulses are simulated in the form of the time-dependent Gaussian-enveloped electric field $\mathbf{E}(t)$, expressed as

$$\mathbf{E} = \sum_{i}^{n} \mathbf{E}_0 \exp\left(-\frac{4\ln 2(t - i \cdot \Delta t)^2}{\sigma^2}\right) \cdot \cos(2\pi\omega(t - i \cdot \Delta t)), \qquad (2)$$

where $\mathbf{E}_0$ is the peak electric field, $\sigma$ is the full width at half maximum, $\Delta t$ is pulse interval, and $\omega$ is frequency. For most calculations here, we use $\mathbf{E}_0$ of 0.4 V/Å, $\sigma$ of 1 ps, $\Delta t$ of 1 ps, and $\omega$ of 1 THz simulations.

The scheme of the electric enthalpy function is employed to determine the response to finite electric fields[70, 71]

$$F(\mathbf{R}, \mathcal{E}) = E_{KS}^0(\mathbf{R}) - \mathbf{P}(\mathbf{R}) \cdot \mathbf{E}, \tag{3}$$

where $E_{KS}^0(\mathbf{R})$ is the zero-field ground-state Kohn-sham energy at the coordinates $\mathbf{R}$, and $\mathbf{P}$ is the polarization. $\mathbf{E}$ is a time-dependent Gaussian-enveloped electric field as Eq. (2). In the presence of an applied electric field, the equilibrium coordinates that minimize the electric enthalpy function should satisfy the force-balance equation

$$-\frac{dE_{KS}^0}{dR} + Z^0 \cdot \mathrm{E} = 0, \tag{4}$$

where $Z^0$ is the zero-field Born effective change tensor calculated by density functional perturbation theory[72]. Note that the Born effective change tensors for AB stacking and BA stacking are different though they are very similar. The electric field is applied as the force $Z(t) \cdot \mathrm{E}(t)$ added in Langevin's equation[70, 71]. Such a scheme has been shown to provide good accuracy for electric-field-related physical responses in ferroelectric and multiferroic compounds[70, 73].

The phonon dispersion is calculated using the frozen-phonon method[74], and the phonon frequencies and normal modes are obtained via the PHONOPY software[75]. The amplitude $Q_\alpha$ of normal mode $\alpha$ is calculated according to $U_j = \sum_\alpha Q_\alpha/\sqrt{m_j} \cdot q_j^\alpha$, where $U_j$ is the displacement of the $j$th atom, $m_j$ is the mass of this atom, and $q_j^\alpha$ is the corresponding component of the normal-mode eigenvector[76]. The units of Å for displacement, amu for masses, and $\text{Å}\sqrt{\text{amu}}$ for the normal mode amplitudes $Q_\alpha$ are used.

## Acknowledgments

The authors thank the National Key R&D Program of China (Grants No. 2022YFB3807601 and No. 2020YFA0711504), the National Natural Science Foundation of China (Grants No. 12274201, No. 52232001, No. 51721001, No. 52003117), the Program for Innovative Talents and Entrepreneur in Jiangsu (JSSCTD202101). We are grateful to the HPCC resources of Nanjing University for the calculations.